# Layer-dependent magnetism and spin fluctuations in atomically thin van der Waals magnet CrPS$_4$

Mengqi Huang,[1] Jazmine C. Green,[2] Jingcheng Zhou,[1] Violet Williams,[3] Senlei Li,[1] Hanyi Lu,[1] Dziga Djugba,[1] Hailong Wang,[4,5] Benedetta Flebus,[3] Ni Ni,[2] and Chunhui Rita Du[1,4,5,*]

[1]Department of Physics, University of California, San Diego, La Jolla, California 92093, USA
[2]Department of Physics and Astronomy, University of California, Los Angeles, California 90095, USA
[3]Department of Physics, Boston College, Chestnut Hill, Massachusetts 02467, USA
[4]Center for Memory and Recording Research, University of California, San Diego, La Jolla, California 92093, USA
[5]School of Physics, Georgia Institute of Technology, Atlanta, Georgia 30332, USA

*Correspondence to: c1du@physics.ucsd.edu

**Keywords:** quantum sensing, nitrogen-vacancy magnetometry, magnetic two-dimensional materials, spin fluctuations, tunneling magnetoresistance.

**Abstract:** van der Waals (vdW) magnets, an emerging family of two-dimensional (2D) materials, have received tremendous attention due to their rich fundamental physics and significant potential for cutting-edge technological applications. In contrast to the conventional bulk counterparts, vdW magnets exhibit significant tunability of local material properties, such as stacking engineered interlayer coupling and layer-number dependent magnetic and electronic interactions, which promise to deliver previously unavailable merits to develop multifunctional microelectronic devices. As a further ingredient of this emerging topic, here we report nanoscale quantum sensing and imaging of atomically thin vdW magnet chromium thiophosphate CrPS$_4$, revealing its characteristic layer-dependent 2D static magnetism and dynamic spin fluctuations. We also show a large tunneling magnetoresistance in CrPS$_4$-based spin filter vdW heterostructures. The excellent material stability, robust strategy against environmental degradation, in combination with tailored magnetic properties highlight the potential of CrPS$_4$ in developing state-of-the-art 2D spintronic devices for next-generation information technologies.

Recent discovery of van der Waals (vdW) magnets has inspired a vast research interest in exploring transformative two-dimensional (2D) spintronic devices capable of delivering novel functionalities, such as ultra-high densities, improved compatibility to device integration, and tailored heterostructure engineering, to next-generation information sciences and technologies.[1,2] To date, $CrX_3$ (X = I, Cl, Br),[3–9] $Fe_3GeTe_2$,[10] $Cr_2Ge_2Te_6$,[11] and $MnBi_2Te_4 (Bi_2Te_3)_n$ [12,13] have been prominent candidates of this family, which exhibit a range of exotic material properties such as layer-dependent magnetic and electronic interactions,[3,14,15] stacking-induced moiré magnetism,[16–19] and unconventional magnetic tunneling effects [20–22] that are currently under intensive investigation. Despite the enormous technological promise and rich fundamental physics, implementation of 2D spintronic devices incorporating vdW magnets for practical applications remains at its infancy. An apparent challenge arises from the materials' ultrahigh sensitivity to air and light exposure,[11,23] which prohibits a broad range of their applications in the ambient environment.

Recently, the emergence of a more robust magnetic vdW material chromium thiophosphate $CrPS_4$ with improved chemical stability provides an appealing platform to tackle the above challenges, expanding the available material scope for 2D spintronic research.[24–27] Here we report nanoscale quantum sensing and imaging of atomically thin $CrPS_4$ crystals. Taking advantage of nitrogen-vacancy (NV) nanomagnetometry techniques,[28–32] we investigate layer-dependent (anti)ferromagnetism and spin fluctuations in $CrPS_4$ nanoflakes. Notably, robust spin fluctuations are observed in both odd and even layer $CrPS_4$ nanoflakes and the obtained magnetic susceptibility monotonically increases with reducing layer numbers. We also prepared $CrPS_4$-based vdW spin filter devices and observed a large tunneling magnetoresistance up to 100% below the magnetic phase transition temperature. Our results highlight the significant potential of $CrPS_4$ for developing robust, air-stable vdW spin logic devices for practical applications. The demonstrated NV quantum metrology techniques could be readily applied to a broad family of vdW crystals, opening a new avenue for probing the spin-related phenomena in emergent quantum states of matter.

We first briefly review the crystal structure and magnetic properties of the vdW material in the current study. $CrPS_4$ has a monoclinic crystal symmetry where each Cr atom is surrounded by six S atoms forming a slightly distorted octahedron $CrS_6$ in a 2D rectangular lattice as shown in Fig. 1a.[24,25] The one-dimensional chains of the chromium octahedra arranged along the *a*-axis are interconnected by P atoms. Below the Néel temperature of 38 K,[26] $CrPS_4$ bulk behaves as a typical A-type antiferromagnet in the magnetic ground state, where magnetic moments carried by $Cr^{3+}$ in individual monolayers align parallel with each other along the *c*-axis direction and any two neighboring magnetic monolayers are antiferromagnetically coupled (Fig. 1a). By applying an external magnetic field $B_{ext}$ along the *c*-axis direction, $CrPS_4$ will be driven to the canted antiferromagnetic phase at spin-flop transition fields around $\pm 7$ kG, and eventually enters the forced ferromagnetic state around 80 kG (Section S1, Supporting Information).[26]

To evaluate the alternating antiferromagnetic and ferromagnetic like behaviors in even and odd layer $CrPS_4$ crystals, we utilize NV wide-field magnetometry techniques [28–31] to image spatially resolved static magnetization and spin fluctuations at the nanoscale. An NV center consists of a nitrogen atom adjacent to a carbon atom vacancy in one of the nearest neighboring sites of a diamond crystal lattice.[32,33] The negatively charged NV state has an $S = 1$ electron spin and serves as a three-level quantum system for implementing qubit-based nanomagnetometry. Figure 1b shows the schematic of our measurement platform, where a (001)-oriented diamond sample containing shallowly implanted NV ensembles is used for wide-field imaging measurements. Atomically thin $CrPS_4$ flakes were first exfoliated on an oxidized Si substrate as shown by an optical microscope image in Fig. 1c. The layer number of individual sample areas is determined by thickness-dependent optical contrast and confirmed by atomic force microscopy characterizations (Section S1, Supporting Information). Exfoliated $CrPS_4$ flakes were subsequently transferred onto the diamond sample for NV magnetometry measurements which will be discussed in detail below.

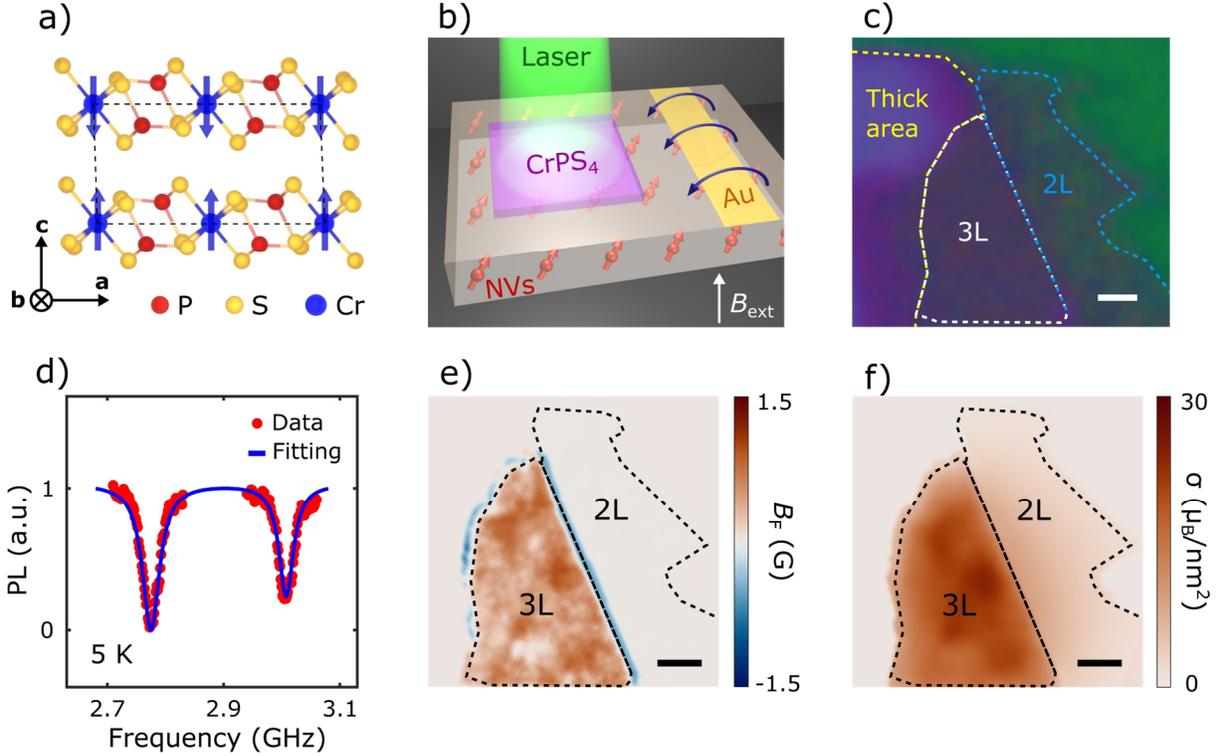

**Figure 1.** Quantum sensing of atomically thin $CrPS_4$ using NV wide-field magnetometry techniques. a) Crystal and magnetic structures of $CrPS_4$. The blue, yellow, and red balls represent Cr, S, and P atoms, respectively. Blue arrows denote local magnetic moments carried by Cr atoms in two neighboring magnetic monolayers of $CrPS_4$. b) Schematic of NV wide-field magnetometry platform in which a $CrPS_4$ nanoflake is transferred onto a diamond substrate containing shallowly implanted NV centers. c) Optical microscope image of an atomically thin $CrPS_4$ crystal with characterized layer numbers for individual sample areas. The crystal is exfoliated onto an oxidized silicon substrate. d) A typical set of optically detected magnetic resonance spectrum showing the Zeeman splitting of NV spin energy levels. e-f) Wide-field imaging of magnetic stray field $B_F$ (e) and reconstructed magnetization $\sigma$ (f) of the prepared $CrPS_4$ flake at 5 K. The dashed lines outline the boundary of the bilayer and trilayer $CrPS_4$ flakes. The scale bar is 2 μm in Figs. 1c, 1e, and 1f.

We first utilize NV centers to investigate the layer-dependent static magnetization of the $CrPS_4$ sample. d.c. wide-field magnetometry utilizes the Zeeman effect [32,33] of NV ensembles to spatially resolve magnetic stray fields emanating from proximate $CrPS_4$ samples. Figure 1d presents a typical set of NV electron spin resonance (ESR) spectrum recorded at a single camera pixel over the $CrPS_4$ sample at 5 K. In the current work, a small external magnetic field $B_{ext}$ of ~70

G is applied along the out-of-plane (OOP) direction of the sample. NV ensembles in the diamond substrate have four possible spin orientations with a mirror symmetry axis along the OOP direction. Considering the spontaneous perpendicular anisotropy of $CrPS_4$, the measured NV ESR spectrum shows four-fold degeneracy with only one pair of split NV spin energy levels (Fig. 1d), from which the magnitude of the OOP stray field $B_F$ exclusively arising from the $CrPS_4$ sample can be measured (Section S2, Supporting Information).[34] By measuring the field-induced Zeeman splitting at every pixel of the captured image, we are able to obtain a magnetic stray field $B_F$ map of the $CrPS_4$ sample as shown in Fig. 1e. Notably, magnetic stray field predominantly emerges from the trilayer $CrPS_4$ flake area while $B_F$ arising from the bilayer $CrPS_4$ exhibits a vanishingly small value. The observed strong even-odd contrast is attributed to the layer-number determined (un)compensated magnetization of atomically thin $CrPS_4$ crystals. In the presented NV wide-field magnetometry measurement platform, the vertical distance between NV spin sensors and the sample surface is ~10 nm, much larger than the thickness of $CrPS_4$ flakes. Thus, magnetic stray fields generated by individual ferromagnetic monolayers of an even layer $CrPS_4$ crystal nearly compensate at the NV sites while an odd layer $CrPS_4$ sample could produce a significantly pronounced stray field due to the uncompensated net magnetization. Note that the sign of the measured stray field reverses across the boundary of the odd layer $CrPS_4$ flake owing to the characteristic dipole nature. Through established reverse-propagation protocols,[7,16] spatially resolved magnetization patterns of the $CrPS_4$ sample can be reconstructed from the stray field image as presented in Fig. 1f. The trilayer $CrPS_4$ flake shows robust magnetization with a spatially averaged value of ~20 $\mu_B/nm^2$, in qualitative agreement with results reported in previous studies.[26] It is instructive to note that the observed spatial variations of local $CrPS_4$ magnetization could result from inhomogeneities, magnetic domains, and defects.[25]

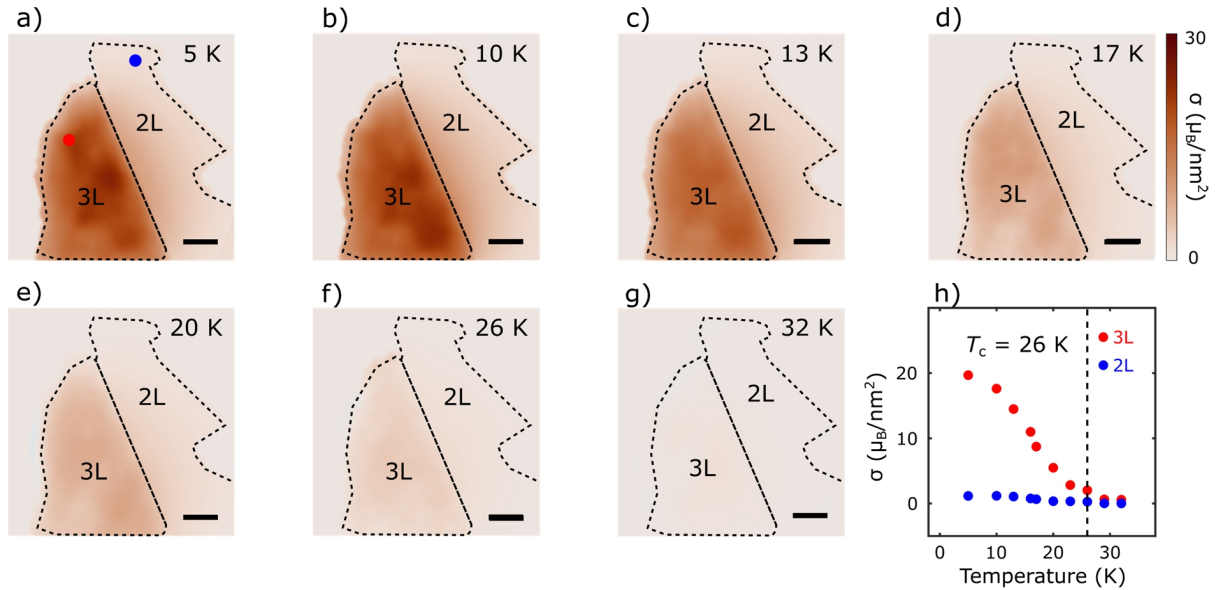

**Figure 2.** Temperature dependent magnetization maps of a $CrPS_4$ flake probed by NV magnetometry. a-g) Magnetization maps of the prepared $CrPS_4$ nanoflakes measured with a perpendicular magnetic field $B_{ext}$ = 70 G at temperatures of 5 K (a), 10 K (b), 13 K (c), 17 K (d), 20 K (e), 26 K (f), and 32 K (g), respectively. The black dashed lines outline the boundary of the bilayer and trilayer $CrPS_4$ flakes of interest, and the scale bar is 2 μm. h) Temperature dependence of magnetization measured at two local sample sites of the bilayer and trilayer $CrPS_4$ flakes.

To investigate the second-order magnetic phase transition of the sample, we performed systematic NV wide-field magnetometry measurements to visualize the evolution of the static CrPS$_4$ magnetization patterns across the Néel temperature. Figures 2a-2g present the reconstructed magnetization maps of the CrPS$_4$ sample measured from 5 K to 32 K. In the low temperature regime ($T \leq 10$ K), the exfoliated CrPS$_4$ sample exhibits robust magnetization in the trilayer flake area, as shown in Figs. 2a and 2b. The measured uncompensated magnetic moment decreases with increasing temperature due to thermally induced spin fluctuations (Figs. 2c-2e). When approaching the Néel temperature of CrPS$_4$ where the energy scale of thermal spin fluctuations becomes comparable to the intralayer ferromagnetic exchange interaction, long-range magnetic order sustained by the intrinsic magnetocrystalline anisotropy starts to collapse, accompanied by the significant decrease of the CrPS$_4$ magnetization (Fig. 2f). At $T = 32$ K above the Néel temperature, no clear magnetic features are observed over the entire CrPS$_4$ flake area as shown in Fig. 2g. Figure 2h summarizes the temperature dependence of the measured magnetic moment at two local sample sites of the trilayer and bilayer CrPS$_4$ flakes. Clear second-order magnetic phase transition is observed with a Néel temperature of ~26 K for the trilayer crystal (Section S3, Supporting Information).

So far, we have shown the strong even-odd contrast of static magnetization of atomically thin CrPS$_4$ crystals. Next, we use wide-field NV spin relaxometry method [30,35–39] to probe intrinsic spin fluctuations in the 2D vdW magnet of interest. Spin fluctuations in an antiferromagnetically correlated system are driven by its time dependent spin density distribution, for example, due to the dynamic imbalance in the thermal occupation of the magnon bands with opposite chiralities.[30] While a fully compensated antiferromagnet exhibits zero net static magnetization, the spin-spin correlation-induced time dependent fluctuations of the average spin density do not vanish.[30] As qubit-based magnetometers, NV centers with excellent quantum coherence are ideally posed to investigate local spin fluctuations in vdW CrPS$_4$ nanoflakes that are challenging to access by conventional magnetometry techniques. The NV spin relaxometry measurements take advantage of the dipole-dipole interaction between spin fluctuations in a magnetic sample and proximal NV centers. Emanating fluctuating magnetic fields at the NV ESR frequencies will accelerate NV spin transitions from the m$_s$ = 0 to m$_s$ = $\pm$1 states, leading to enhancement of the corresponding NV spin relaxation rates [30,40] (Section S2, Supporting Information). By measuring the spin-dependent NV photoluminescence, the occupation probabilities of NV spin states can be quantitatively obtained, allowing for extraction of the NV spin relaxation rate which is proportional to the magnitude of the local fluctuating magnetic fields transverse to the NV axis.

Figures 3a-3f present a series of spin fluctuation induced NV spin relaxation rate $\Gamma_M$ maps of the CrPS$_4$ sample measured at temperatures from 13 K to 44 K. It is evident that robust magnetic fluctuations with a strong temperature dependence are observed in both bilayer and trilayer CrPS$_4$ crystals as expected from the time dependent spin-spin correlations of the (un)compensated magnetic density.[30,40] To further reveal the layer-number dependent spin fluctuations, Fig. 3g summarizes temperature dependence of spatially averaged NV spin relaxation rates for CrPS$_4$ crystals with different layer numbers. Note that the tetralayer and six-layer CrPS$_4$ samples were separately prepared on other diamond substrates for NV measurements (Section S3, Supporting Information). One can see that the measured NV spin relaxation rates consistently show an

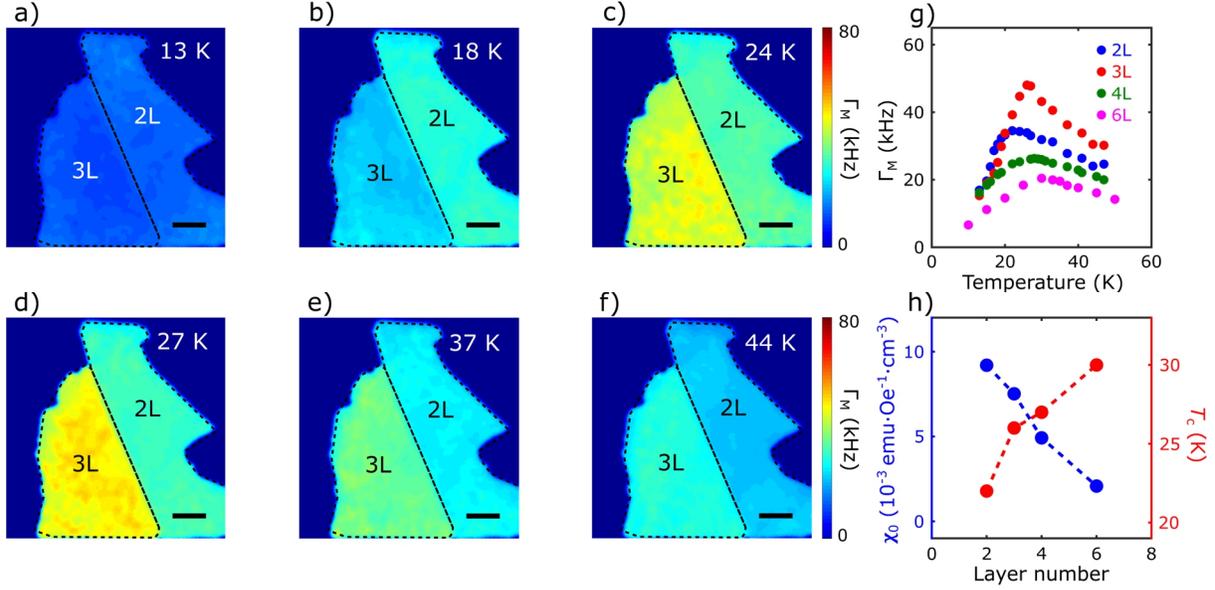

**Figure 3.** Quantum imaging of spin fluctuations in an atomically thin CrPS$_4$ flake. a-f) NV spin relaxation maps of the prepared bilayer and trilayer CrPS$_4$ flakes measured at temperatures of 13 K (a), 18 K (b), 24 K (c), 27 K (d), 37 K (e), and 44 K (f), respectively. The NV ESR frequency is set to be 2.78 GHz in these measurements with an external perpendicular magnetic field $B_{ext}$ of 70 G. The dashed lines outline the boundary of the bilayer and trilayer CrPS$_4$ flakes of interest, and the scale bar is 2 μm. g) Temperature dependence of spatially averaged NV spin relaxation rate $\Gamma_M$ for bilayer, trilayer, tetralayer, and sixlayer CrPS$_4$ crystals, showing peak values around the corresponding magnetic phase transition temperatures. h) Layer number dependence of magnetic critical temperature $T_c$ and longitudinal magnetic susceptibility $\chi_0$ of atomically thin CrPS$_4$ crystals measured at 20 K.

enhancement around the magnetic phase transition of CrPS$_4$ nanoflakes, which is attributed to the divergent magnetic susceptibility $\chi_0$ around the quantum phase transition point.[24] When $T$ is above 30 K, moderate spin fluctuations remain present in CrPS$_4$ due to the finite spin-spin correlations in the paramagnetic state.[41] The obtained Néel temperatures of individual CrPS$_4$ nanoflakes qualitatively increases with the layer number as shown in Fig. 3h, in agreement with previous studies on vdW magnets [10,11]. Fundamentally, the measured NV spin relaxation is mainly driven by the longitudinal spin fluctuations of CrPS$_4$, which is directly related to the static longitudinal magnetic susceptibility $\chi_0$ of the sample.[30,40,42] By invoking a theoretical model developed in Ref. 30, longitudinal magnetic susceptibility of atomically thin CrPS$_4$ crystals can be quantitatively measured as presented in Fig. 3h (Section S4, Supporting Information). When $T = 20$ K, $\chi_0$ of six-layer CrPS$_4$ is estimated to be 2.3 × 10$^{-3}$ emu·Oe$^{-1}$·cm$^{-3}$. As the layer number decreases, magnetic order of CrPS$_4$ crystals becomes less robust and more vulnerable to external perturbations, resulting in a significantly enhanced magnetic susceptibility. It is worth mentioning that the NV spin relaxation rate measured around Néel temperatures first increases with the layer-number of the vdW crystals, reaching a peak value for trilayer CrPS$_4$. Further increase of the sample thickness results in significant decay of $\Gamma_M$. The observed experimental feature can be explained by the two competing factors on the measured NV spin relaxation rates: sample thickness and intrinsic magnetic susceptibility $\chi_0$ of CrPS$_4$ crystals. In the low thickness regime (≤ 3 layer), sample thickness is the dominant factor that determines the magnitude of spin fluctuations in CrPS$_4$ while magnetic susceptibility plays a more prominent role in the large sample thickness regime (> 3 layer).

In comparison with conventional vdW magnets, a distinct advantage of $CrPS_4$ results from its robust material stability against air exposure, providing an attractive platform for developing practical spintronic applications.[24,25] Next, we report our initial efforts along this direction by showing a large tunneling magnetoresistance in $CrPS_4$-based vdW heterostructures. Figure 4a illustrates the device structure for electrical transport measurements, in which an atomically thin $CrPS_4$ layer acts as a spin-filter tunnel barrier sandwiched by graphite electrical contacts. Figure 4b shows an optical microscope image of a prepared tunnel junction device, where blue and black dashed lines outline the boundaries of the tetralayer $CrPS_4$ and graphite flakes, respectively (Section S1, Supporting Information). Considering the overlap between the top and bottom graphite electrodes, the effective tunnel junction area is estimated to be ~4 μm². Figure 4c plots the characteristic nonlinear dependence of tunneling current ($I_t$) on the applied d.c. bias voltage ($V_{bias}$). For the atomically thin CrPS$_4$ tunnel layer used in the current study, each magnetic monolayer acts as an independent spin filter that is oppositely aligned in series. The double spin-filtering effect-induced tunneling magnetoresistance arises when the magnetizations of neighboring CrPS$_4$ monolayers are switched from an antiparallel to parallel state. Figure 4d shows the tunneling electrical conductance $G$ as a function of the applied perpendicular magnetic field $B_{ext}$ at 10 K with a d.c. bias voltage of 300 mV. In the low field regime ($B_{ext} \leq 4$ kG), the tetralayer CrPS$_4$ stays in the antiferromagnetic ground state, showing a relatively small $G$. When $B_{ext}$ exceeds ~ 7 kG, field-induced spin-flop transition drives CrPS$_4$ to the canted antiferromagnetic phase accompanied by a sudden jump of the tunneling conductance. Further increasing $B_{ext}$, the measured tunneling conductance gradually saturates to a plateau value as CrPS$_4$ evolves to the fully aligned ferromagnetic state at ~80 kG. We have proposed a theoretical model to rationalize the observed spin-dependent electronic tunneling effect in tetralayer CrPS$_4$ (Fig. 4d) (Section S5, Supporting Information). The tunneling magnetoresistance defined as $(G_p - G_{ap})/G_{ap}$ reaches ~100 %, where $G_{ap}$ and $G_p$ are the tunneling conductance of the junction device in the antiferromagnetic and ferromagnetic states of CrPS$_4$, respectively.

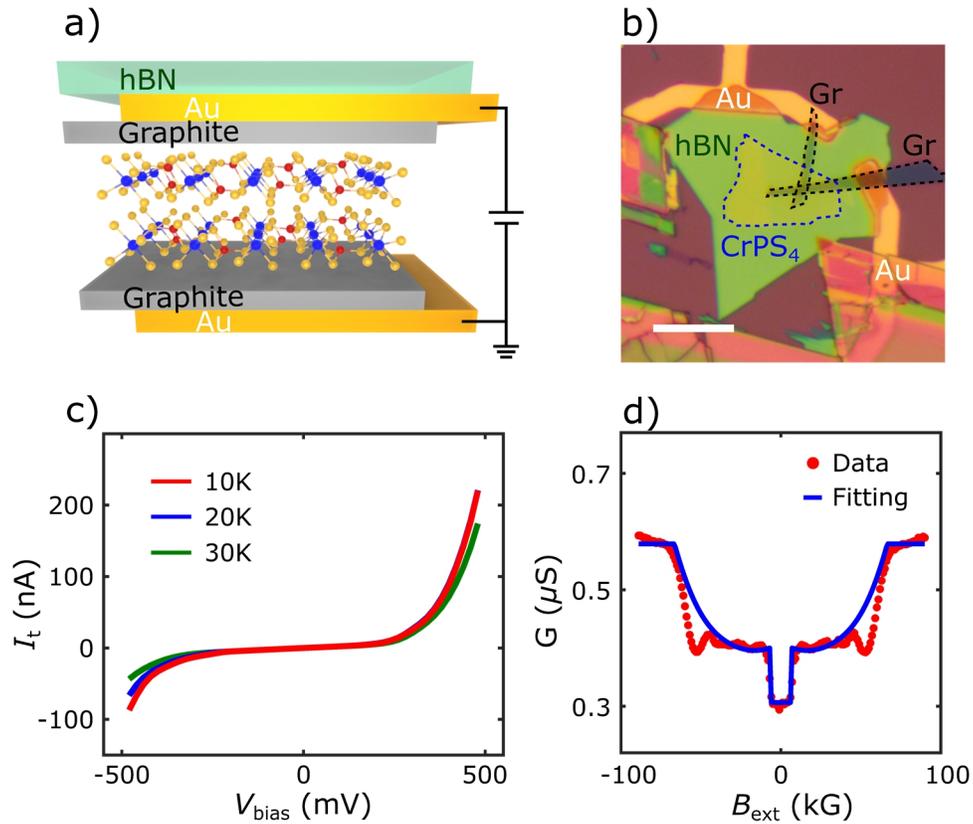

**Figure 4.** Tunneling magnetoresistance measured in a $CrPS_4$-based vdW junction device. a) Schematic of a vdW spin-filter tunnel junction device consisting of an atomically thin $CrPS_4$ layer sandwiched by two graphite electrodes. b) Optical microscope image of a prepared $CrPS_4$-based vdW tunnel junction device. The blue and black dashed lines outline the boundaries of the tetralayer $CrPS_4$ and graphite flakes, respectively, and the scale bar is 20 μm. c) d.c. bias voltage dependence of the tunneling current of a $CrPS_4$-based vdW junction device measured at 10 K, 20 K, and 30 K in absence of an external magnetic field. d) Tunneling conductance $G$ as a function of an out-of-plane magnetic field $B_{ext}$, in qualitative agreement with a theoretical model. The experimental results were measured at 10 K with application of a d.c. bias voltage of 300 mV.

In summary, we have demonstrated NV wide-field imaging of layer dependent static magnetization and dynamic spin fluctuations in atomically thin CrPS$_4$ crystals. By fabricating CrPS$_4$-based vdW junction devices, we further report a large tunneling magnetoresistance, which is fundamentally related to different magnetic states of the tetralayer CrPS$_4$ tunneling layer. The observed prominent magneto-tunneling effect, the field controllable magnetic orders, together with the robust chemical stability make CrPS$_4$ a promising material candidate to develop functional 2D spintronic devices for practical applications. Our results also highlight the potential of NV centers in probing nanoscale spin related phenomena in low-dimensional vdW materials, opening the possibility of engineering NV-vdW-magnet-based hybrid quantum systems [43] for developing next-generation information technologies.

**Author contributions:** M.H. performed the NV measurements and analyzed the data with H. L and H. W. J. Z., S. L., and D. D. prepared the devices. S. L. performed the electrical transport measurements. J. G. and N. N. provided CrPS$_4$ bulk crystals. V. W. and B. F. provided theoretical input on the tunneling magnetoresistance measured in CrPS$_4$-based spin filter vdW heterostructures. C. R. D. supervised the project.

**Notes:** The authors declare no financial interest.

**Acknowledgements**: M. H., H. L., H. W., and C. R. D. were supported by the Air Force Office of Scientific Research under award No. FA9550-20-1-0319 and its Young Investigator Program under award No. FA9550-21-1-0125. J. Z. and D. D. acknowledged the support from U. S. National Science Foundation (NSF) under award No. DMR-2046227. S. L. acknowledged the support from U.S. Department of Energy (DOE), Office of Science, Basic Energy Sciences (BES), under award No. DE-SC0022946. Work at UCLA was supported by DOE, Office of Science, Office of Basic Energy Sciences under Award No. DE-SC0021117. B. F. was supported by the NSF under Grant No. NSF DMR-2144086.